\documentclass[manuscript]{acmart}
\pdfoutput=1

\usepackage{tabularx}
\usepackage{color, colortbl}
\definecolor{shade}{gray}{0.95}

\AtBeginDocument{%
  \providecommand\BibTeX{{%
    \normalfont B\kern-0.5em{\scshape i\kern-0.25em b}\kern-0.8em\TeX}}}

\copyrightyear{2023}
\acmYear{2023}
\setcopyright{acmcopyright}\acmConference[]{}
\acmBooktitle{}
\acmPrice{15.00}
\acmDOI{}
\acmISBN{}

\begin{document}

\title[Evaluation of Sketch-Based and Semantic-Based Modalities for Mockup Generation]{Evaluation of Sketch-Based and Semantic-Based Modalities for Mockup Generation}

\author{Tommaso Cal{\`o}}
\orcid{0000-0002-3200-2348}
\affiliation{%
  \institution{Politecnico di Torino}
  \streetaddress{Corso Duca degli Abruzzi, 24}
  \city{Torino}
  \postcode{10129}
  \country{Italy}}
\email{tommaso.calo@polito.it}

\author{Luigi De Russis}
\orcid{0000-0001-7647-6652}
\affiliation{%
  \institution{Politecnico di Torino}
  \streetaddress{Corso Duca degli Abruzzi, 24}
  \city{Torino}
  \postcode{10129}
  \country{Italy}}
\email{luigi.derussis@polito.it}

\renewcommand{\shortauthors}{Cal\`o and De Russis}

\begin{abstract}
Design mockups are essential instruments for visualizing and testing design ideas. However, the process of generating mockups can be time-consuming and challenging for designers. In this article, we present and evaluate two different modalities for generating mockup ideas to support designers in their work: (1) a sketch-based approach to generate mockups based on hand-drawn sketches, and (2) a semantic-based approach to generate interfaces based on a set of predefined design elements. To evaluate the effectiveness of these two approaches, we conducted a series of experiments with 13 participants in which we asked them to generate mockups using each modality. Our results show that sketch-based generation was more intuitive and expressive, while semantic-based generative AI obtained better results in terms of quality and fidelity. Both methods can be valuable tools for UI designers looking to increase their creativity and efficiency.
\end{abstract}

\ccsdesc[500]{Human-centered computing~Graphical user interfaces}
\ccsdesc[300]{Human-centered computing~Interface design prototyping}
\ccsdesc[500]{Computing methodologies~Machine learning}
\ccsdesc[500]{Computing methodologies~Computer vision}

\keywords{machine learning, web elements, user interface, convolutional neural network}


\maketitle

\section{Introduction And Motivation}
User interface designers face daily challenges in creating designs that are effective, usable, and innovative. They often draw inspiration from existing design samples to come up with new ideas \cite{arr}. There are two main types of resources that support this process of finding inspiration \cite{insp}. The first type are design gallery platforms, such as Dribbble \cite{dribble} and Behance \cite{Behance}, which allow designers to browse through a collection of designs and find examples that are interesting or useful for their work. The second type are design inspirational tools that suggest examples based on certain types of design input, such as a sketch or an existing design, using algorithms to determine image similarity \cite{guifetch, exa7,dtour,rewire}.\\
While these approaches can be helpful in finding inspiration, they have limitations. Browsing through design galleries can be overwhelming and lead to a shift in design ideas away from the original focus, while relying too much on examples with similar styles can lead to design fixation and hinder the originality of the work \cite{fix, ari}. 
To find a balance between targeted and serendipitous inspiration, Mozaffari et al. \cite{gansp} proposed a style-based generative adversarial network (StyleGAN) trained on a large dataset of existing interface designs. It generates a diverse yet focused set of examples based on a preliminary design input. While it can generate a diverse range of interface mockups, the user control is limited to injecting style into the latent space of the model. In other words, the user can specify certain style attributes (such as color and texture) for the model to incorporate, but has little control over the specific layout or design elements of the mockup. A better approach would allow designers to explicitly specify the desired layout and other design elements, rather than relying on the model to randomly generate these elements based on style injection. This would enable a more targeted and precise design process, as designers can focus on specific aspects of the mockup rather than being limited to injecting style into the model.
In this paper, we aim to achieve this goal by exploring two distinct AI-assisted modalities for mockup generation: one based on sketches, and the other based on semantic-colored drawings. Sketch-based mockup generation involves the use of hand-drawn sketches to represent the desired design. Semantic-colored drawing-based mockup generation, instead, involves the use of detailed, colored drawings that convey specific design elements. 
The choice of these two modalities was based on a careful review of the literature, indicating that sketching and semantic drawing are commonly used in the design process due to their ability to balance precision and speed \cite{li2023gligen,park2019SPADE,calo}. Sketch-based mockup generation allows for quick exploration of ideas while semantic-colored drawing-based mockup generation provides more accuracy in representing the final product. To achieve this, we used the Pix2Pix \cite{pix2pix} model for sketch-based mockup generation and the SPADE \cite{spade} model for semantic-colored drawing-based mockup generation. The Pix2Pix model is well suited for sketch-based mockup generation because it is a conditional GAN that has been proven to be effective in generating high-quality images from sketches \cite{color}. Similarly, the SPADE model is well suited for semantic colored drawing-based mockup generation as it is a GAN that is trained to generate images from semantic maps, it allows for more control over specific design elements, which is crucial for conveying accurate representations of the final product.
We evaluate the two modalities of mockup generation in terms of their expressivity, time demand, ease, and intuitiveness. By analyzing the advantages and disadvantages of both sketch-based and semantic-colored drawing-based mockup generation, we aim to empower designers to select the most appropriate modality for their particular design project and boost their creative potential.

\section{Related Works}
Inspiration has long been of interest to researchers in the field of psychology. Thrash et al. \cite{trash,thrash2014psychology} were among the first to systematically study inspiration, identifying three categories: motivation (i.e., goal-oriented self-initiation), evocation (i.e., impulsive reactions to stimuli), and transcendence (i.e., feeling of gaining superior, more elegant or novel ideas than those generated willfully). These studies have generally focused on how designers access and utilize existing design artifacts. For instance, Eckert and Stacey \cite{eckert2000sources} found that knitwear designers used a variety of sources for inspiration, including artifacts with intriguing shapes, patterns, and colors, as well as their own previous work. Researchers have also explored how industrial and user interaction designers are inspired by existing design artifacts. Bonnardel \cite{bonnardel1999creativity} found that in the context of product design, ``the emergence of new ideas results from analogy-making.'' Herring et al. \cite{herring2009getting} conducted in-depth interviews with web, graphic, and product designers and identified common approaches and challenges they faced in retrieving, storing, and disseminating design examples. Gonçalves et al. \cite{gonccalves2014inspires} surveyed students and industrial designers to understand their sources and methods of inspiration and found that professional industrial designers used a wider variety of approaches compared to students. The literature has also identified several issues with common inspirational methods, including the risk of ``design fixation'' from exposure to a homogenous set of examples and the impact of timing on the quantity and quality of ideas.

In recent years, Artificial Intelligence (AI) has emerged as a potential tool for supporting and enhancing the creative inspiration process \cite{shneiderman2022human}. AI systems can be trained to generate ideas and outputs based on a set of rules or guidelines, enabling the efficient production of a wide range of options, and providing a variety of applications and systems to support professionals in various visual art fields, such as graphic design \cite{contgrad}, UI design \cite{inproceedings}, webtoon \cite{wetoon}, digital art \cite{drawing8}, and new media art \cite{initim}. To ensure a balance between human control and computer automation \cite{hcai}, some AI systems have employed a ``human-in-the-loop'' approach, which allows for collaboration between humans and AI. Overall, AI has been seen as a helpful tool in the design process, aiding in ideation and generating a variety of output efficiently.

Our work compares two modalities, sketches and semantic-colored drawings, to generate user interface mockups and build upon previous knowledge in several ways.
First, the use of sketches as a modality builds upon previous research on the use of visual representations in the design process \cite{sket}. Studies have shown that sketches can be used to quickly generate and communicate design ideas, allowing designers to explore a wide range of possibilities in a relatively short amount of time. Additionally, sketches are often used as a way to capture informal, early-stage design ideas, which can then be refined and developed further \cite{910894}. Second, the use of semantic-colored drawings \cite{sem0,zhu2020sean}, where each color represents a UI element category, allows for a clear representation of the design elements and their relationships. The two modalities work together towards the goal of allowing designers to quickly generate and communicate ideas using sketches, while also leveraging the power of color coding to represent the design elements and their relationships. Both modalities combine the benefits of computer automation and human control \cite{shneiderman2022human} enabling the efficient production of a wide range of options while maintaining high fidelity to the original design idea.

\begin{figure*}[htbp]
  \centering
 \includegraphics[width=0.6\linewidth]{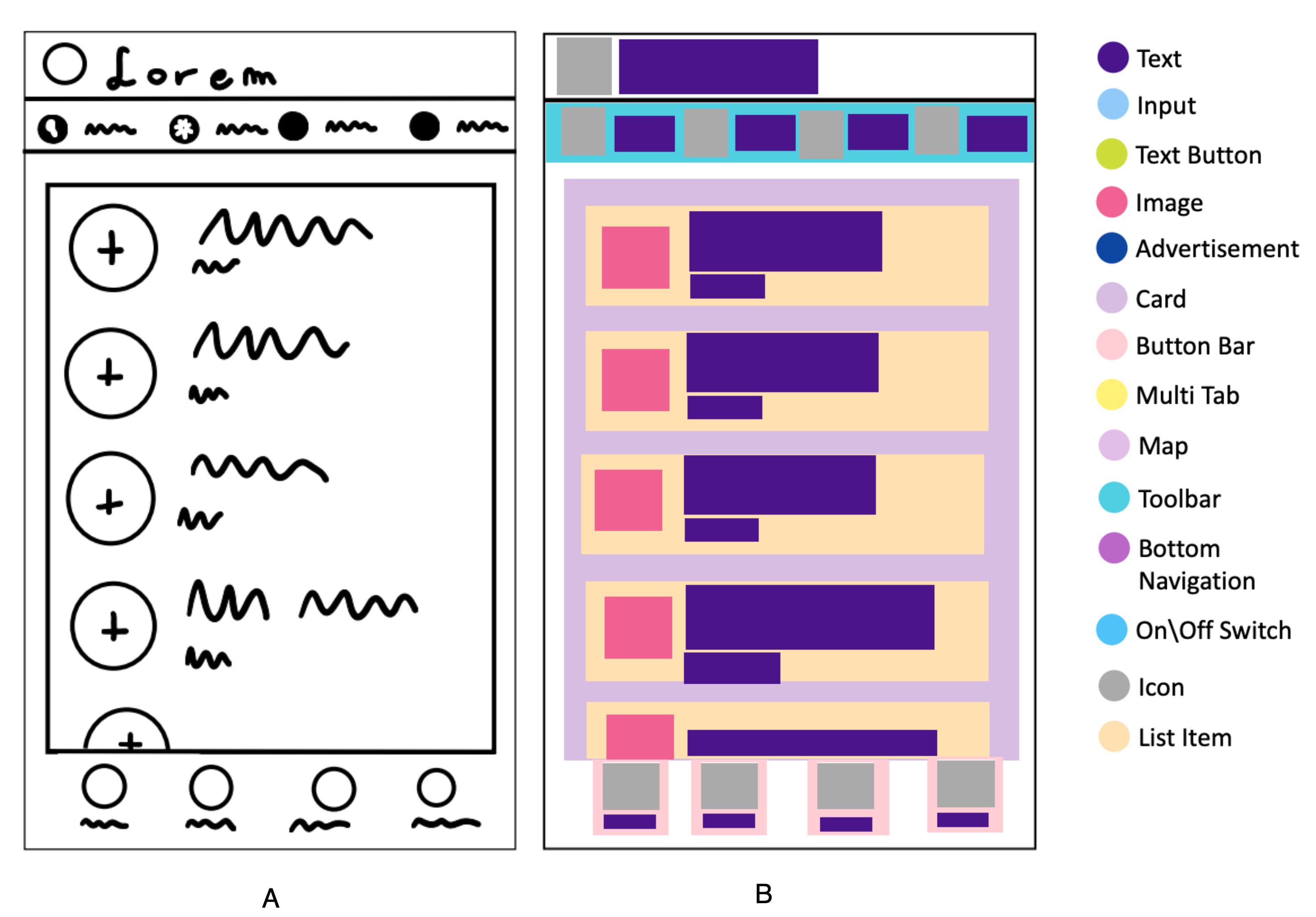}
  \caption{An example of the same user interface represented in the sketch-based and semantic-based modalities. On the left, we can see a hand-drawn sketch of the user interface (A), with various elements such as buttons, text fields, and icons depicted in a freeform manner. On the right, we can see the same user interface represented in the semantic-based modality (B). These elements are filled with different colors, with each color representing a specific design element such as buttons, text fields, and icons. }
  \Description{The figure shows the skeleton of a mobile user interface with titles, navigation, and a list of items as both a hand-drawn sketch and a semantic-colored representation, where each design element is represented by a different color.}
  \label{fig:experiment}
\end{figure*}

\section{Study: Evaluating Generation Modalities}
This study aimed to evaluate the semantic drawing-to-mockup and sketch-to-mockup modalities in terms of time demand, ease, creative expressivity, and intuitiveness. The study was conducted with 13 human-computer interaction students from the first-year master's program in computer engineering. The participants were asked to use both modalities to create a mockup of a mobile application and then rate their experience with each modality on various dimensions. The results of the study provide insights into the strengths and weaknesses of each modality and can inform the design and development of mobile applications in the future.

\subsection{Participants}
The study was conducted with 13 participants who gave their consent to participate in the study. All participants were studying Human-Computer Interaction (HCI) as part of their curriculum. It is important to note that all participants in the study had received training in UI design and programming as part of their curriculum, indicating that they had at least some level of experience in this area. The sample consisted of 2 female and 11 male participants, with an age range of 23 to 25 years old. Given that the sample is relatively small, it can be considered an exploratory study. Future studies should replicate and expand the study on a larger sample to generalize the findings. The study aimed to be a first step to understanding the differences between the two modalities.

\subsection{Procedure}
Participants were asked to complete a two-phase task requiring them to design a mobile user interface by sketching and semantically drawing it. The task was designed to evaluate the participants' abilities in terms of time demand (time to complete), expressiveness (how well the modality conveyed their idea), intuitiveness (in understanding the modality), and ease of use (of the modality).

In the first phase, participants were sent a document containing instructions on how to complete the task. They were asked to design a mobile user interface and sketch in a given frame. While participants were allowed to use their preferred digital drawing tool in the study to ensure a more natural and intuitive design process, future studies may consider restricting the use of digital drawing tools to a single option to better control for potential variations in user experience. After sketching, participants were instructed to semantically draw the user interface. The order of the tasks was chosen based on the natural progression of a typical design process in which initial, unstructured ideas and concepts are often followed by more concrete iterations. This allowed us to evaluate the strengths and weaknesses of each modality in facilitating the design process, rather than comparing the effectiveness of one modality over the other. An example from the experiment, showcasing the same user interface represented in both sketch and semantic-drawing form, can be seen in Figure~\ref{fig:experiment}.
Once the participants had completed the task, they were asked to rate their experience in terms of the above dimensions on a Likert scale. Participants were also asked to provide motivation for their ratings.

In the second phase, the participants' drawings were processed by a network (see ``implementation'' below), and users were asked two additional questions about the quality of the obtained results and how the results respected the imagined user interface. The participants were asked to rate the quality of the results obtained from the network on a Likert scale, and to provide feedback on how well the results corresponded to their imagined user interface.

\begin{figure*}[htbp]
  \centering
 \includegraphics[width=0.8\linewidth]{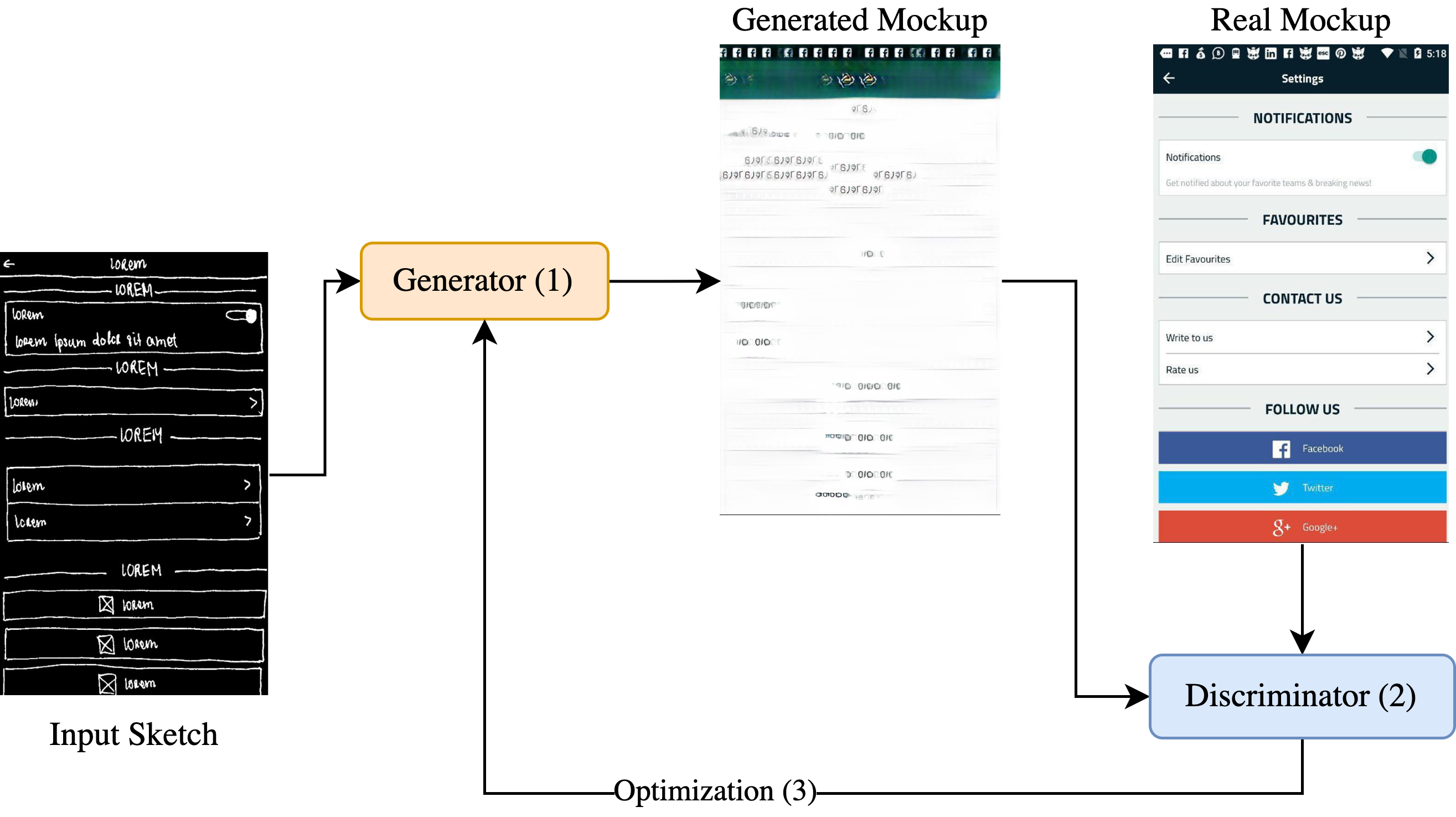}
  \caption{Illustration of how Pix2pix \cite{pix2pix} translates between sketches and real mockups domains. A generator is trained to generate mockups from sketches (1) and a discriminator is trained simultaneously to distinguish between the real and generated mockups (2). A Reconstruction loss, in addition, measures how close the generated mockup is to the real target mockup (3). The only difference with SPADE \cite{spade} is that in the latter a  SPatially-Adaptive Normalization module is added to optimize the translation of semantic maps.}
  \Description{An architectural diagram: the Pix2pix algorithm starts from the input sketch, pass it through a generator to generate a mockup. The generated mockup and the corresponding real user interface go to a discriminator that provides optimization parameters to the generator, to improve the classification results.}
  \label{fig:framework}
\end{figure*}

\subsection{Implementation}
To translate sketches and semantic colored drawing to mockups, we use two deep learning models: Pix2pix~\cite{pix2pix} and SPADE~\cite{spade} trained on RICO~\cite{Deka:2017:Rico} dataset. Pix2pix is the model we used to generate realistic mockups from sketches, while SPADE is the model we used to generate mockups from semantic-colored drawings, an illustration of the framework used to generate mockups can be found in Figure \ref{fig:framework}. The trained models are available for download at \url{HIDDENLINK}\footnote{Temporary link for preserving the anonimity during the peer-review phase.} for the reproducibility of results.

\subsubsection{Dataset}
In this study, we used the RICO dataset \cite{Deka:2017:Rico} for both semantic drawing-to-mockup and sketch-to-mockup modalities. The RICO dataset is a publicly available dataset of hand-drawn sketches and corresponding semantic representations of mobile app user interfaces. It is composed of a total of approximately 20k sketches and 60k semantic representations. The dataset is designed to be a benchmark for evaluating the performance of deep learning models for mobile app user interface generation. To ensure the quality of the data, we corrected the colors in the semantic representation that were corrupted and released the corrected version of the dataset on the same GitHub repository. This was done to ensure that the results of the study were not affected by any errors or inconsistencies in the data.
We found the RICO dataset to be particularly useful for our study as it provided a large and diverse set of sketches and semantic representations of mobile app user interfaces. 

\subsubsection{Sketch-to-mockup}
Pix2pix is a model that uses a conditional generative adversarial network (cGAN) to generate images. It consists of two networks: a generator and a discriminator. In our case, the generator takes as input a sketch and generates the corresponding mockup, while the discriminator receives both the generated image and the target image and tries to distinguish between the two. The two networks are trained simultaneously, with the generator trying to fool the discriminator and the discriminator trying to classify the images correctly.

\subsubsection{Semantic drawing-to-mockup}
SPADE (Semantic Image Synthesis with SPatially-Adaptive Normalization) is a model that uses a style-based generator architecture to generate images from semantic colored drawings. It consists of a generator network that takes as input a semantic-colored drawing and generates a corresponding image, and a discriminator network that receives both the generated image and the target image and tries to distinguish between the two. The generator network includes SPatially-Adaptive Normalization (SPADE) layers that adapt the normalization parameters of the generator based on the input semantic map, optimizing the image synthesis for semantic map inputs. This is achieved through the use of a separate normalization network, which takes the input image and a set of spatial locations as input and produces the normalization parameters for each location. 

\section{Preliminary Results}
\begin{table}[h]
    \renewcommand*{\arraystretch}{1.25}
    \small
    \caption{Comparison of sketch-to-mockup and semantic drawing-to-mockup modalities on intuitiveness, ease, time demand, and expressiveness, as well as the quality and fidelity of the generated mockups, on a 1 to 5 Likert scale.}
    \label{tab:my_label}
    \centering
    \begin{tabular}{ cccc } 
        \toprule
        & \textbf{Sketch-to-mockup} & \textbf{Semantic drawing-to-mockup} \\
        \midrule
        \textbf{Intuitiveness} & 4.54 $\pm$ 0.78 & 3.15 $\pm$ 1.28  \\
        \rowcolor{shade}
        \textbf{Ease of use} & 3.38 $\pm$ 1.04 & 3.69 $\pm$ 1.25 \\
        \textbf{Time Demand} & 3.54 $\pm$ 1.13 & 3.77 $\pm$ 1.09 \\
        \rowcolor{shade}
        \textbf{Expressiveness} & 4.15 $\pm$ 0.55 & 3.38 $\pm$ 1.12 \\
        \hline
        \textbf{Quality} & 1.50 $\pm$ 0.58 & 2.75 $\pm$ 0.50  \\
        \rowcolor{shade}
        \textbf{Fidelity} &  3.38 $\pm$ 2.20 & 3.60 $\pm$ 0.55 \\
        \bottomrule
    \end{tabular}
\end{table}

The study compares the sketch and semantic drawing modalities by gathering participant feedback on their level of intuitiveness, ease of use, and time demand during the generation process, as well as the quality and fidelity of the generated mockups. The results of the survey provide insight into how participants perceive these modalities and their relative strengths and weaknesses. Numerical results are reported in Table \ref{tab:my_label}.

\paragraph{Intuitiveness:}
The results of the comparison between the sketch and semantic drawing modalities to input an AI for generating mockups show that the sketch generation score for intuitiveness is generally higher than the semantic generation score. Most participants (11 out of 13) gave a score of 5 for the sketch generation, indicating that they found it fully intuitive. In contrast, only two participants gave a score of 5 for the semantic generation, with the majority of participants (8 out of 13) giving a score of 3 or lower.
One of the main motivations for participants finding the sketch generation more intuitive is that it allows them to ``show others what you have in mind'' and ``represent at a higher level each part of the layout'' (participant 1, 6). Another motivation is that it is easier to ``imagine the layout'' (participant 9) and ``pour the idea in mind into drawings'' (participant 10) with the sketch generation.
However, some participants had issues with the semantic generation, finding it ``difficult due to the rules about the colors'' (participant 4) and ``not so clear how it can be useful'' (participant 1). Participant 12 also found semantic generation less intuitive stating that it is not simple for everyone to understand.
Conversely, some participants found the semantic generation to be more intuitive, with participant 10 stating that it is more intuitive as it has a ``map (indication) about element types'' which makes things ``much easier to demonstrate graphically''.

\paragraph{Ease of use:}
The results of the comparison between the sketch and semantic drawing modalities for ease of use indicate that there is no statistical significance between the mean ease of use of the two methods. Most participants (7 out of 13) gave a score of 3 or lower for the sketch generation, indicating that they found it to be difficult to use.  One of the main motivations for participants finding sketch generation difficult to use is that it requires ``drawing skills and tools'' (participant 9) and ``some drawing skills are needed to convey information in a reasonable and clear way'' (participant 1). Additionally, some participants found it difficult to use the mouse to draw (participants 3, 4, 11) and some participants found it difficult to use because they had to use different devices to complete the task (participant 13).
Participants found the semantic modality difficult to use because it "needs to remember the colors according to the legend" (participant 9), while on the other hand, some participants found it easy to use because it ``only requires squares and rectangles'' (participant 4) and is ``just a mechanical exercise'' (participant 3). Participant 10 also found the semantic generation easy to use because it is ``much more descriptive'' and ``easier regarding positioning the items'' as compared to sketch generation.

\paragraph{Time demands:}
When it comes to time demand, participants had mixed opinions on the time efficiency of sketch and semantic drawing modalities.
The main motivation for participants finding the sketch generation quicker is that it allows them to represent the layout at a higher level and it is easier to imagine the layout and that ``only took me some minutes to complete'' (participant 5). However, some participants experienced challenges with the sketch generation, citing issues with tools such as ``difficulty drawing with a mouse'' (participant 10) and ``difficulty in getting a decent result'' (participant 3). On the other hand, the semantic generation was perceived as less time demanding by some participants, with participant 8 stating that it is ``the quickest one because it’s similar to sketch generation but with even fewer details to represent'' and participant 3 finding it ``fairly quick to draw what I was thinking in the semantic way''. However, other participants found it to be more time demanding, with participant 1 stating that they ``spent very much time in finding a software to draw colored squares'' and participant 12 stating that ``it can take a long time because there is often a difference between ideas and reality.''

\paragraph{Expressivity:}
The results of the comparison between the sketch and semantic drawing modalities for expressivity show that the expressivity score for the sketch generation is generally higher than the semantic generation score. Most participants (10 out of 13) gave a score of 4 or 5 for the sketch generation, indicating that they found it to be fully expressive. The majority of participants (7 out of 13) gave a score of 3 or lower for the semantic generation, indicating that they found it to be less expressive.
Some of the main motivations for participants finding the sketch generation to be more expressive is that it ``makes it easiest to describe how you want the final result to look like'' (participant 3) and ``irons out some ambiguity'' (participant 5). Participants also found it to be expressive because it allows them to ``draw almost everything'' (participant 5) and ``the alignments and the position of the items are easier to express'' (participant 6).
On the other hand, some participants found the semantic generation to be less expressive because it ``just conveys information with colored and filled squares that don't reflect very well the actual elements'' (participant 1) and it is ``too restrictive'' (participant 4). Participants also found it difficult to ``define the role of each component'' (participant 7). Overall, the results suggest that participants generally find the sketch generation to be more expressive for inputting mockups into an AI, but some participants also found the semantic generation to be useful, particularly for its ability to convey information through color-name association.

\begin{figure*}[htbp]
  \centering
 \includegraphics[width=0.9\linewidth]{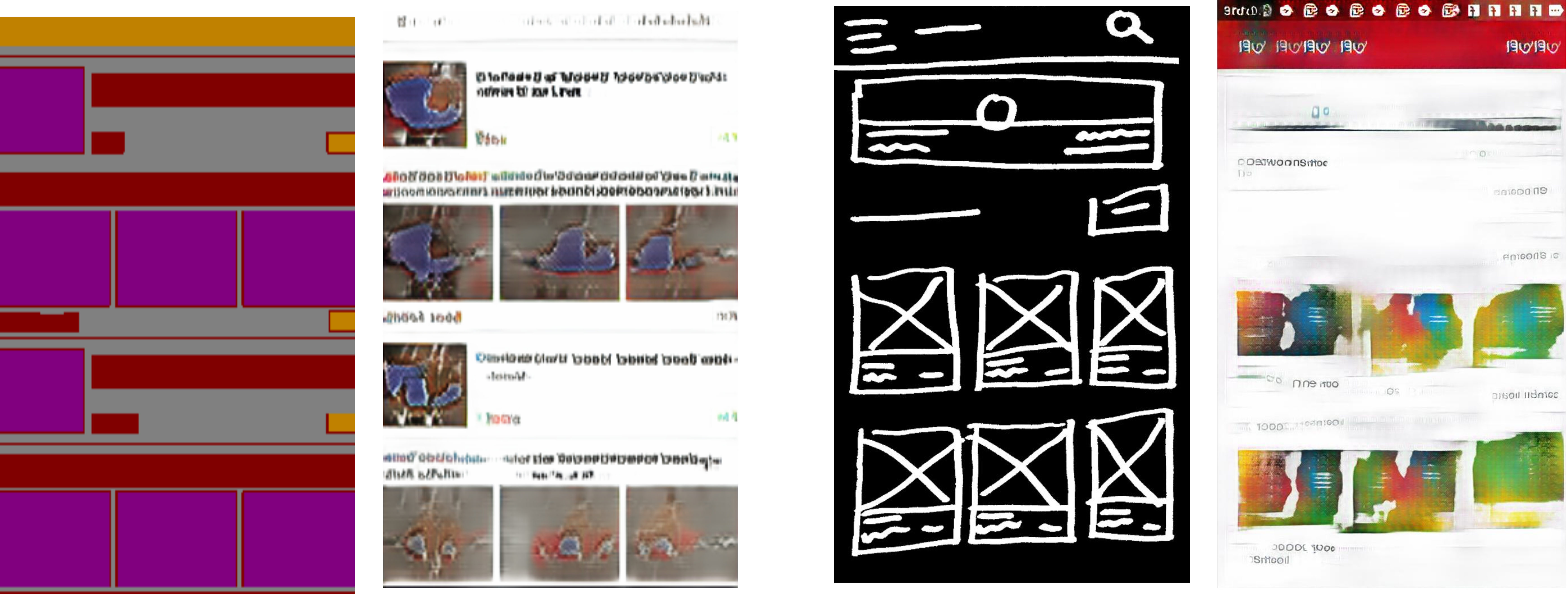}
  \caption{An example of two mockups generated through semantic-drawing modality (left) and sketch modality (right). The image demonstrates the improved resulting quality and reduced presence of artifacts achieved through the use of semantic-drawing modality.}
  \Description{The generated mobile mockups along with their input drawings in the two modalities. }
  \label{fig:ex}
\end{figure*}

\paragraph{Quality and fidelity of the generated mockups:}
Results indicate that the quality of the generated mockup for the sketch-to-mockup modality is higher than for the semantic drawing-to-mockup modality with a mean of 2.75 and 1.50 respectively. In terms of fidelity to the original idea, the sketch-to-mockup modality also scores higher with a mean of 3.60 compared to 3.38 for the semantic drawing-to-mockup modality. However, this difference is not as large as for the quality of the generated mockup but still statistically significant. In summary, the results indicate that the sketch-to-mockup modality generally performs better in terms of quality of the generated mockup and fidelity to the original idea compared to the semantic drawing-to-mockup modality, an example is reported in Figure \ref{fig:ex}. However, it is worth noting that these results should be taken with caution as it is based on a small sample size, and it would be beneficial to have more data and a larger sample size to make more robust conclusion.

\paragraph{Additional Comments}
In analyzing the additional comments, it is clear that there are different preferences for the best method of inputting AI to generate mockups. Participant 7 suggests that the sketch generation modality is the best, but that the problem of describing components' function can be solved by adding a color convention to the sketch generation. This suggests that the participant values the intuitiveness and expressivity of the sketch generation modality, but also recognizes the benefits of using a semantic approach for describing the function of components.
Participant 8 also prefers the sketch generation modality, citing its ability to describe the interface in more detail than the semantic method. 
Participant 11 suggests that an hybrid solution between the sketch and semantic modality may be a good option. Participant 11 recognizes the limitations of both modalities, specifically the lack of semantic meaning in the sketch generation and the difficulty of representing specific shapes in the semantic method, and suggests that combining them may provide the best solution.
Participant 12 believes that the best method depends on the context. They also believe that there is no one method that is better than others, but they are complementary and useful to provide a complete description.

\section{Conclusions and Future Work}
The paper evaluated the sketch and semantic drawing modalities by gathering user feedback on their level of intuitiveness, ease of use, and time demand during the generation process, as well as the quality and fidelity of the generated mockups. 

The results of the study indicate that users generally found the sketch generation to be more intuitive and expressive compared to the semantic generation. The study provides insight into how users perceive these modalities and their relative strengths and weaknesses and can help inform future design decisions for AI mockup generation tools. Many users have suggested that a possible way to optimize the interaction between the user and the system would be to combine the two approaches, taking advantage of the strengths of each modality while minimizing their weaknesses. For example, by using sketching to quickly and intuitively create a rough prototype, and semantic drawing to add more detailed and expressive design elements, designers would have a more flexible and powerful tool for creating their mockups. Additionally, this approach could help designers to express their ideas and designs more easily and accurately, which may result in a better final product. 

In conclusion, a mixed approach that combines sketching and semantic drawing could be a promising solution for generating mockups of mobile applications and additional research should explore this possibility.





\bibliographystyle{ACM-Reference-Format}
\bibliography{style-sketch-code.bib}


\end{document}